\documentclass[a4paper,11pt]{article}
\pdfoutput=1 % if your are submitting a pdflatex (i.e. if you have
\usepackage{jcappub} % for details on the use of the package, please
\usepackage[T1]{fontenc} % if needed
\usepackage{aas_macros}

\def\d {\rm{d}}
\newcommand{\D} {{\tilde{\rm{D}}}}

\newcommand{\<}{\langle}
\renewcommand{\>}{\rangle}

\title{\boldmath Gravitational edge mode powers galaxy flat rotation curves}

 \author{Obinna Umeh}
\affiliation[a]{Institute of Cosmology \& Gravitation, University of Portsmouth, Portsmouth PO1 3FX, United Kingdom}
\affiliation[b]{Department of Physics, University of the Western Cape, Cape Town 7535, South Africa}
\affiliation[c]{Hierarchical Intelligence Lab, Divine Estate, Charlisco Phase 2, Warri, Delta State, Nigeria}

\emailAdd{obinna.umeh@port.ac.uk}
\abstract{
The point-particle approximation is foundational to modelling clustering of matter in the universe, but is fundamentally inconsistent within General Relativity due to associated spacetime singularities. This bottleneck has historically restricted the study of matter clustering to linear scales. We resolve this by utilising the recent observation that a matter horizon precedes the formation of caustics in expanding spacetimes. This allows for the isolation of singularities via spacetime surgery. By glueing distinct spacetime sheets related by a discrete transformation across the shared boundary, we derive a covariant backreaction term that contributes to the effective energy-momentum tensor. Crucially, we identify this backreaction contribution with gravitational edge modes—physical degrees of freedom residing on boundaries that arise from the breaking of diffeomorphism invariance. These gravitational edge modes modify local particle trajectories, naturally producing flat galaxy rotation curves in the outskirts without invoking dark matter particles. Our framework thus demonstrates that gravitational edge modes can act as effective dark matter, offering a first-principles alternative to particle dark matter for explaining galactic dynamics.

}

\begin{document}
\maketitle
\flushbottom

\section{Introduction}
\label{sec:intro}

The standard model of cosmology, while remarkably successful on large scales, exhibits persistent and conceptually troubling shortcomings on small scales. At the level of statistical descriptors, such as the matter power spectrum and higher-order N-point correlation functions, perturbative treatments break down as one approaches nonlinear, small-scale regimes, leading to divergences or uncontrolled sensitivities to ultraviolet physics ~\cite{Bernardeau:2001qr,Umeh:2015gza,Umeh:2016nuh,Jolicoeur:2017eyi,Jolicoeur:2017nyt,Jolicoeur:2018blf,Koyama:2018ttg,Clarkson:2018dwn,Umeh:2019qyd,Umeh:2019jqg,Maartens:2020jzf}. These issues are not merely technical but signal a deeper inconsistency in how matter is modeled: the idealization of matter as a collection of point particles neglects internal structure, finite size effects, and the gravitational backreaction that becomes increasingly important in dense environments~\cite{Umeh:2026ajv,Umeh:2026mac}. A similar pathology appears in the prediction of cosmological observables. Quantities such as the luminosity distance and angular diameter distance, when computed in inhomogeneous spacetimes, develop divergences tied to unresolved small-scale structure~\cite{Umeh:2022hab,Umeh:2022prn,Umeh:2022kqs}. This suggests that the standard smoothing procedures and effective descriptions are insufficient, and that key physical contributions are being systematically neglected.

Compounding these issues is that the standard model description of structure formation is fundamentally entangled to the dark matter particle and Point Particle Approximation (PPA)~\cite{Hahn:2015sia, Adamek:2016zes}. 
 The breakdown of PPA is sometimes attributed to the nonlinear nature of the equations of General Relativity(GR)~\cite{Poisson:2011nh, Senovilla:2014kua}, but the problem is more fundamental; it is due to the finite nature of the geodesic on curved spacetime for some class of initial conditions~\cite{Umeh:2023lbc,Umeh:2026ajv}.
We show that the breakdown of PPA is a manifestation of the discrete nature of spacetime(nature allows separation of scales). PPA assumes that geodesics propagate on a fixed background spacetime; the influence of the particle on the spacetime itself is neglected~\cite{Poisson:2003nc}.  We approach this using the technique of Matched Asymptotic Expansions (MAE) in cosmology~\cite{Poisson:2003nc,Goldberg:2016lcq} to capture the impact of the particle, which develops a matter horizon in regions where particle gravity exactly cancels Hubble flow~\cite{Ellis:2010fr,Umeh:2026ajv}.  MAE allows the use of PPA on scales where it is valid while capturing the backreaction effect through the boundary term. We apply manifold surgery(cut and glue) at the level of the action rather than to the equation of motion using the variational principle, which ensures a consistent covariant treatment of the variational principle, boundary conditions and covariant interpretation of backreaction effects of spacetime on the particle trajectory~\cite{Umeh:2023lbc}.
%Finally, we describe how the backreaction effects manifest as dark matter on small scales through its impacts on the galaxy rotation curve.
 Crucially, we identify these boundary contributions as gravitational edge modes, a well-studied concept in quantum gravity~\cite{Takayanagi:2019tvn,Donnelly:2020xgu,Donnelly:2022kfs}. These are physical degrees of freedom that reside on the boundaries separating local structures from the cosmological spacetime. 
 As an example, we show that these gravitational edge modes are all we need to explain the galaxy flat rotation curves.

This paper is structured as follows: in section \ref{sec:structure}, we review structure formation within the standard model of cosmology and provide a consistent definition of the matter horizon in general relativity. We describe hierarchical structure formation within cosmological zoom-in perturbation theory in section \ref{sec:Hierrachial} and conclude in section \ref{sec:conc}.
We use the  Planck CMB constraint on cosmological parameters for quantitative estimate: $h = 0.674$ for the dimensionless Hubble parameter, $\Omega_{\rm b} = 0.0493$ for baryon density parameter,  $\Omega_{\rm{cdm}} = 0.264$ for the dark matter density parameter,  $\Omega_{\rm m} = \Omega_{\rm{cdm}} + \Omega_{\rm b}$ for the matter density parameter,  $n_{\rm s} = 0.9608$ for spectral index,  and  $A_{\rm s} = 2.198 \times 10^{-9}$ for the amplitude of the primordial curvature perturbation~\cite{Planck:2018vyg}. 
The small English alphabet from $a-e$ denotes the full spacetime indices, while $i$ and $j$ denote the spatial indices.  The capital English alphabet from $A-E$ denotes tetrad indices on the screen space.

\section{Structure formation and astrophysical matter horizon}\label{sec:structure}

\subsection{Review of structure formation within the standard model}

\begin{figure}[h]
 \includegraphics[width=120mm,height=70mm]{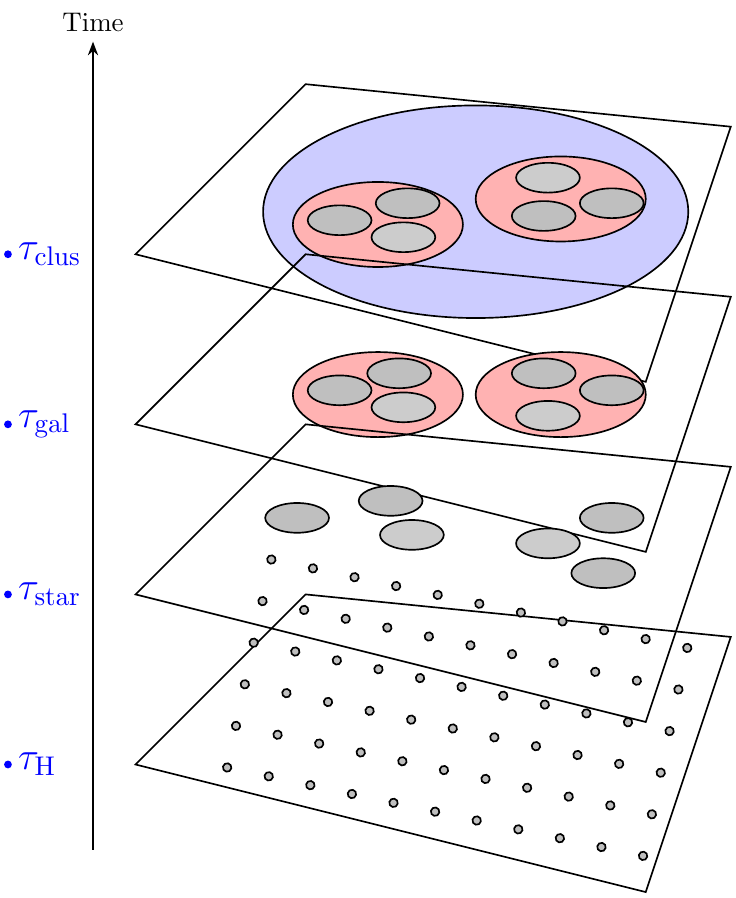}
 \caption{This is an illustration of how light elements(particles), through gravitational collapse instability,  form massive particles after some time has elapsed. The left vertical axis indicates the flow of proper time,  The left vertical line has crucial time scales:  $\tau_{ \rm{H}}$, $\tau_{ \rm{star}}$,  $\tau_{ \rm{gal}}$ and $\tau_{ \rm{clus}}$ which denote maximal hypersurfaces for the hierrachy of mass of corresponding to gravitational bound sysytem, for example: starting from the Hydrogen atom,  to star, galaxy and cluster.  
}
 \label{fig:obs_multi_scale}
\end{figure}

The canonical treatment of structure formation within the standard model of cosmology begins with the linearised evolution of the  density fluctuation $  \delta(\mathbf{x}, t)  = ({\rho - \bar{\rho}})/{\bar{\rho}}$ on an FLRW background spacetime 
\begin{equation}\label{eq:growth_eqn}
    \ddot{\delta} +2H{\delta }+\left(\frac{c_{s}^{2}k^{2}}{a^{2}}-4\pi G {\rho }\right)\delta =0
\end{equation}
where $c_{s}^{2}$  is the square of the speed of sound, $H$ is the Hubble rate.  This equation well-posed in the regime where $\delta \ll 1$, In this regime, it describes a balance between gravity (the term $4\pi G \bar{\rho} \delta$,  Hubble drag $(2H\dot{\delta})$ and the fluid internal pressure. It admits both growing and decaying solutions
$
\delta(\mathbf{k},t) \sim a(t) $  in the matter domination ($a \sim t^{2/3}$, $H = 2/3t$), hence $\delta(\mathbf{k},t)  = a^{-3/2}
$. The small-scale perturbations (high $k$) have higher pressure resistance and therefore oscillate or wash out. while the large-scale perturbations (low $k$) are dominated by gravity and grow. Because the Jeans length (the threshold for collapse, $k_{J}={\sqrt{4\pi G\rho }}/{c_{s}}$) was smaller in the past, smaller density fluctuations were the first to win the battle against pressure and collapse.
The small-scale structures form first (i.e., stars and dwarf galaxies). These structures then merge and are pulled together by the slower-growing, larger-scale perturbations and form galaxies, then clusters of galaxies. See figure \ref{fig:obs_multi_scale} for a schematic illustration of how the process happened. 

However, this story is incomplete; the gravitationally bound systems we observe today, such as galaxies and clusters, have $\delta_{\text{today}} \gg 1$. The standard model of structure formation in the universe explains this by invoking the {spherical collapse model}, where a spherical shell of comoving radius $r$ containing mean overdensity $\delta_i$ evolves as a closed sub-universe \cite{Peebles:1980lssu.book.....P}.
$  \ddot{R} \;=\; -{G\,M(<R)}/{R^2},
$
The equation of motion for the spherical shell
predicts a turnaround at $t_{\star}$ when $\dot{R} = 0$, after which the shell formally collapses in finite time $t_{\mathrm{collapse}} = 2t_{\star}$. The standard model abruptly terminates the collapse trajectory by imposing virialisation as a boundary condition, asserting that the final radius must satisfy $ R_{\mathrm{vir}}  = {R_{\mathrm{ta}}}/{2},
$
derived from the virial theorem $2K + W = 0$. 
The standard model provides no continuous trajectory from $t_{\star}$ to the virialised end-state since virialisation is a statement about a time-averaged equilibrium state, not really a dynamical mechanism.

The absence of this dynamical bridge is not a minor technical gap. It indicates a deeper technical loophole that the standard model of structure formation cannot predict how matter is distributed on small scales.  This gap has restricted small-scale clustering analysis to phenological or empirical models, such as the halo occupation distribution (HOD) \cite{Sheth:1999mn,Sheth:2001dp,Cooray:2002dia} and subhalo abundance matching (SHAM)\cite{Masaki:2022zte},  often combined with N-body simulations \cite{Nishimichi:2011jm,Chaves-Montero:2015iga}. These models, while powerful, remain heuristic in nature.
Consisitent clustering analysis is only possible  on large scales ($k \lesssim 0.1\;h\,\mathrm{Mpc}^{-1}$), where the perturbation theory is reliable;  on small scales, one-loop corrections in standard perturbation theory (SPT) yield $
    P_{\mathrm{1\text{-}loop}}(k)
 = P_{\mathrm{lin}}(k) + P_{22}(k) + P_{13}(k),
$
where
$
    P_{22} = \int d^3q\; P_{\mathrm{lin}}(q)\,
                   P_{\mathrm{lin}}\!\left(|\mathbf{k} - \mathbf{q}|\right),
$
 which is clearly  UV-divergent. Specifically, in the limit $k\to\infty$: $
    P_{13}(k) \sim -k^2\,\sigma_v^2\,P_{\mathrm{lin}}(k),
    \sigma_v^2 = \frac{1}{6\pi^2}\int_0^\infty P_{\mathrm{lin}}(q)\,dq.$
This integral diverges for blue-tilted or scale-invariant spectra without a UV cutoff. The Effective Field Theory of Large Scale Structure (EFTofLSS; \cite{Baumann:2010tm}) attempts to regularise this by introducing counter-terms,
\begin{equation}
P_{\text{EFT}}(k) \approx P_{\text{lin}}(k) + P_{\mathrm{1\text{-}loop}}(k)- {2 \pi [2 c_s^2(a) + c_{visc}^2(a)] \frac{k^2}{k_{\text{NL}}^2} P_{\text{lin}}(k)} + {C_{\text{stoch}} \frac{k^4}{k_{\text{NL}}^4}}
\end{equation}
%The Pressure is captured by $c_s^2$, The Anisotropic Stress (viscosity) is captured by $c_{visc}^2$.
where $P_{\mathrm{1\text{-}loop}}(k) = P_{22}(k) + P_{13}(k)$.  $c_s^2$ and $c_{visc}^2$ are free parameters which may be related to the speed of sound and anisotropic stress (viscosity).  $C_{\text{stoch}} $ is a stochastic term; it accounts for the fact that the short-distance modes (which are integrated out) are not perfectly correlated with the long-distance modes. 
%At leading order, this scales as $k^4$ for the matter power spectrum. 
The exact nature of these terms is not predicated by the standard model of cosmology.  
Essentially, EFTofLSS does not extend treatment to small scales; it replaces hard momentum cut-offs of SPT with counterterms and
it implicitly assumes separation of scales without establishing that there is separation of scales~\cite{Anastasiou:2022udy}. Note that the power spectrum( or the two-point correlation function) is an observable; therefore, this is equivalent to saying that observables are inconsistent within the standard model of cosmology. 
Furthermore, observables such as the luminosity distance and angular diameter distance become ill-defined in collapse structures
 since null geodesics experience infinite lensing convergence,
$
    \kappa  
    \to \infty
    \quad \text{as } b \to 0,
$
for impact parameter $b \to 0$ on a point mass~\cite{Umeh:2022hab,Umeh:2022kqs,Umeh:2022prn}.
All these failures share a common origin: the treatment of matter within the point particle of approximation of matter in the universe is insufficient on small scales. 
Critically, a point particle has no internal degrees of freedom: no spin, no deformability, no internal energy, this implies that features which are crucial on small scales are missing.

Finally, even if the standard model of cosmology manages to explain all these gaps in structure formation, the fundamental nature of cold dark matter on which it is based remains unresolved. Therefore, any efforts at probing the foundational aspects of cosmology should always be encouraged. 
The motivation for the present work is in accordance with this philosophy.  We extend this programme which was started in ~\cite{Umeh:2026ajv}  by drawing a connection between the boundary contribution, which was described as a backreaction effect in~\cite{Umeh:2026mac}, and the gravitational edge modes \cite{Donnelly:2020xgu}. We provide a systematic and geometrically well-defined framework in which the boundary contribution acts as an effective dark matter that explains the observed flat rotation curves.
These gravitational edge modes arise as physical degrees of freedom associated with the breaking of diffeomorphism invariance at finite boundaries. They naturally encode additional gravitational energy that is not captured by the local bulk stress-energy tensor associated with the standard model matter. Our work links aspects of quantum gravity(gravitational edge modes) with observed galaxy flat rotation curve~\cite{Donnelly:2022kfs}.

\subsection{Point particle approximation and structure formation}

Gravity is the primary driver of the dynamics of large-scale structures of the Universe.  The action of the gravitational field on a manifold, $M$ with metric $g_{ab}$ and the dynamics of matter in $(M,g)$ under the influence of gravity is given by a sum of the actions of the matter field and the geometry 
\begin{eqnarray}\label{eq:action}
S = S_{g} + S_{m} = \int \left[ {L}_{\rm{g}}  + {L}_{\rm{m}}\right]\sqrt{-g} \d^4 {x}\,,
\end{eqnarray}
where $S_{g}  $ is the action for the geometry(gravitational field), the correspoonding lagrangain is given by
${L}_{\rm{g}} = (R-2\Lambda)/2\kappa$, with $\kappa = 8\pi G/c^{4}$, $R$ is the Ricci and $\Lambda$ is the cosmologica constant.   $S_{m} $ is the action of the matter fields. The standard approach is to approximate the matter in the universe with an ensemble of point particles. In this limit, the Lagrangian is a  weighted sum of the point particle action with the Dirac delta function 
\begin{eqnarray}\label{eq:Nbody-Lagrangian}
{L}_{\rm{m}} = \sum_{\ell}S_{\ell}\left(x^a(\tau_{\ell}),{x^a}'(\tau_{\ell}) \right) \frac{\delta^{(4)}\left(x^a - x^a_{\ell}(\tau_{\ell})\right)}{ \sqrt{-g} }\,,
\end{eqnarray}
where $\sqrt{-g(t,x^i)}$ is the square root of the metric tensor determinant, $\delta^{(4)}\left(x^a - x^a_{\ell}(\tau_{\ell})\right)$ is the 4-D Dirac delta function, it is normalised to unity
and   $S_{\ell}(x^a_{\ell},{x^a_{\ell}}')$  is the  action for the $\ell$-th massive particle ~\cite{Adamek:2016zes}
\begin{eqnarray}\label{eq:massive_particle_actionl}
S_{\ell}(x^a_{\ell},{x^a_{\ell}}')  = -m_{\ell} \int_{\tau_i}^{\tau_f}  \sqrt{-g_{ab} \frac{\d x^a_{\ell} }{\d\tau_{\ell}}\frac{\d x^b_{\ell}}{\d\tau_{\ell}} } \d\tau_{\ell} \,,
\end{eqnarray}
where   $\tau_{i}$  and $\tau_{f}$ are the initial and final proper time of the particle, ${u^a_{\ell}}' = {\d x^a_{\ell}}/{\d\tau_{\ell}}$, %${x^a_{\ell} }$ is the trajectory of the ${\ell}$-th particle, 
$x_{\ell}^a(\tau_{\ell})$ is the spacetime trajectory of the massive $\ell$-th particle, $m_{\ell }$ is the rest mass of the ${\ell}$-th particle and $\tau_{\ell}$ is the proper time for the ${\ell}$-th particle.  
 Note $x^a$ is a point in the spacetime, and the delta function is non-zero only when $x^a$ coincides with  $x^a_{\ell}(\tau_{\ell})$: 
Variation of the Einstein-Hilbert action with respect to the metric tensor yields a term which does not vanish at the boundary $\partial M$. Mathematically, one usually fixes the metric at the boundary so that $\delta g_{ab}\big|_{\partial M} =0$(Dirichlet boundary conditions ) or requires that the normal derivative of metric variation vanishes, i.e., $(u^c \nabla_c \delta g_{ab} = 0)$.
Standard cosmology usually assumes that these boundary terms are not there; in this limit, the principle of least action leads to the Einstein field equation
 \begin{eqnarray}
 R_{ab} - \frac{1}{2}g_{ab} R = \kappa T_{ab}\,,% \qquad{\rm{and}}  \qquad\nabla_{b} F^{ab} = J^{a}\,,
\end{eqnarray}
In practice, this assumption has some operational implications:  First, it assumes that the underlying spacetime is either asymptotically flat (Minkowski) or that the universe is closed.  A closed manifold has no boundary $(\partial M = \emptyset)$, hence the boundary term is identically zero. We will come back to this later, but it is important to state this. 
The  total energy-momentum tensor is constructed from the action of a massive particle equation \eqref{eq:massive_particle_actionl}) :  $T_{\ell }^{ab} \equiv -({2}/{\sqrt{-g}})
  ({\delta\left(\sqrt{-g} {L}_{\ell}\right)})/{\delta g_{ab}} $ 
 \begin{eqnarray}\label{eq:EMT_particles}
 T^{ab} = \sum^N_{\ell}  T^{ab}_{\ell}
 = \sum^N_{\ell}  \frac{m_{\ell}}{\sqrt{-g}} \int {\d}\tau\, u^{a}_{\ell} u^{b}_{\ell} \delta^4(x - x_{\ell}(\tau_{\ell})) \,.
\end{eqnarray}
The massive point particle action given in equation \eqref{eq:massive_particle_actionl} is used in cosmology to propagate particles the size of clusters, galaxies, stars, Hydrogen atoms, etc, as test particles on a given background spacetime. The  diffeomorphism invariance implies that the action is invariant $(\delta S = 0)$ under general translation $x^a \to x^a + \xi^a$, hence using $\delta _{\xi }g_{ab }=\nabla _{a }\xi _{b }+\nabla _{b }\xi _{a }$ gives the covariant conservation equation $\nabla _{a }T^{a b }=0$.

We focus on the matter-dominated era, that is when the universe had cooled to about 1 billion Kelvin, leading to the formation of light elements such as hydrogen, helium, and small amounts of lithium and beryllium~\cite{Ellis:2010fr}. This is the regime we can confidently use the point particle action given in equation \eqref{eq:action} without worrying about coupling to other fundamental forces of nature. 
%\(M_{total}=\rho _{b}\times V\approx 1.49\times 10^{53}\text{\ kg}\)\(N=\frac{M_{total}}{m_{p}}\approx \mathbf{8.9\times 10}^{\mathbf{79}}\)
At about the Big Bang Nucleosynthesis (BBN) era, there are approximately $N={M_{\rm{total}}}/{m_{p}}\approx {8.9\times 10}^{\mathbf{79}}$ hydrogen atoms in the universe that were created during this period~\cite{Planck:2018vyg}; it is computationally expensive to track the interaction between these atoms using equation \eqref{eq:EMT_particles}, therefore, we take a fluid approximation by coarse-graining  a system of discrete microscopic particles 
\begin{eqnarray}\label{eq:standardEMT}
T^{ab}_{\rm{fluid}}
&\equiv& \frac{1}{\Delta V}    \sum _{i=1}^{N}  T^{ab}(\gamma_{i}) \Delta \gamma^{i} 
 \xrightarrow{\Delta \gamma_{\pm}\rightarrow 0} 
  \frac{1}{ V}  \int_{\Sigma} T^{ab} \sqrt{h} \d^3\gamma \,,
  \label{eq:fluidEqn}
 \end{eqnarray}
 where $\gamma^{i}$ is the coordinate position of a microscopic particle. In the limit where the particle size is small compared to the size of the universe: $\Delta \gamma^{i}/ \Delta  \ll1 $, we replace the sum with integrals, here  $V = \int \d^{3 } \gamma\sqrt{h}$. 
 For each particle at a point $p \in M$, the tangent space decomposes as $T_p M = \langle u^a_{\ell} \rangle\oplus {\Sigma_p} $ where $\langle u^a_{\ell} \rangle$ is the span, $\Sigma_p$ is the orthogonal complement of $u^a_{\ell}$ and the projected metric tensor  on $\Sigma_p$ is given by $h_{\ell ab}\equiv g_{ab}+ u_{\ell a}u_{\ell b}$. 
 In the fluid limit, we decompose the four velocities of the individual microscopic particles $ u^a_{\ell} $ into macroscopic  and random(thermal) velocity parts:
  $u^a_{\ell} = u^a_{\rm{mac}} + w^a_{\ell}$, where  $u^a_{\rm{mac}} $ is the macroscopic part of the 4-velocity and  $w^a_{\ell}$ is a random or thermal velocity contribution. Just to reduce clutter,  we use $ u^{a}$ in place of $u^a_{\rm{mac}} $ in the rest of the discussion. 
 After coarse-graining,  our spacetime  locally looks like  $
M \simeq \mathbb{R} \times \Sigma,$
where \(\mathbb{R}\) parametrizes the proper time \(\tau\) along the average four velocity $u^a$, and
 $\Sigma$ denotes a hypersurface. 
 
Performing the integration in equation \eqref{eq:fluidEqn},  requires simplification of the following terms
  \begin{eqnarray}\label{eq:quadvelocity}
     \sum_{\ell =1}^{N}  m_{\ell} {u}^a_{ \ell} {u}^b_{ \ell}     &=&    \sum_{\ell =1}^{N}  m_{\ell}
     u^a u^b +       \sum_{\ell =1}^{N} m_{\ell}u^a w^b_{\ell} +      \sum_{\ell =1}^{N} m_{\ell}w^a_{\ell} u^b +      \sum_{\ell =1}^{N} m_{\ell}w^a_{\ell} w^b_{\ell}  \,.
 \end{eqnarray}
Here, the first term denotes the mass-energy density ($\rho_{m}$):
  $\rho_{ m} \equiv \sum_{\ell \in  V} m_{\ell}/{ V}  = M/V_{\pm}$, where $ M =\sum_{\ell \in  V} m_{\ell} $ .
The weighted average of the thermal velocity  fluctuations vanishes: $\sum_{\ell =1}^{N}  m_{\ell} w^a_{\ell}  = 0$ by the definition.
 The fourth term leads to a stress-tensor: 
     $   \mathcal{P}^{ab}  =   \frac{1}{ V} \sum_{\ell =1}^{N} m^{\pm}_{\ell}w^a_{\ell} w^b_{\ell} $, which can be decomposed further  into isotropic and anisotropic parts: $
\mathcal{P}^{ab} = \mathcal{P}_{\pm} h^{ab}_{\pm} + \pi^{ab}_{\pm},$
where $\mathcal{P} = \frac{1}{3} h_{ab} \mathcal{P}^{ab}_{\pm} = \frac{1}{3} \frac{1}{V} \sum_{\ell \in  V} m_{\ell}  h_{ab} w^a_{\ell} w^b_{\ell} $  is the isotropic pressure,  and $
 \mathcal{P}^{\langle ab \rangle}= \big( h^{(a}_{\;c} h^{b)}_{\;d} - \frac{1}{3} h^{ab} h_{cd} \big) \mathcal{P}^{cd} $ is the traceless symmetric projection of the anisotropic stress.  ${h}_{ab}={g}_{ab}+{u}_a {u}_b$ is the metric on the hypersurface orthogonal to $u^a$, 
 Therefore, in the fluid limit, the energy-momentum tensor for an ensemble of particles  is given by
 \begin{eqnarray}\label{eq:emt}
T^{ab}_{\rm{fluid}}=    u^au^b \rho_{m }  + \mathcal{P}  h^{ab} + \mathcal{P}^{\langle ab \rangle} \,.
 \end{eqnarray}
 The pressure contribution can be simplified further by defining an observable-weighted average of the squared random velocity  $\mathcal{P}  = \frac{1}{3} \frac{1}{V \rho_{m}} \sum_{\ell \in  V} m_{\ell}  h_{ab} w^a_{\ell} w^b_{\ell}  \rho_{m}$, The velocity dispersion can now be defined as 
  \begin{eqnarray}
  \langle w^2 \rangle_{\rm{m}} 
 = {\sum m_{ \ell} | w_\ell|^2}/{\sum m_{\ell}}  =  \sigma_{m }^2
 \end{eqnarray}
 Hence,    $\mathcal{P}  = \rho_{ m \pm } \sigma_{1D}^2 $ $\sigma_{m}^2 = 3\sigma_{m 1D}^2$, where $\sigma_{m 1D}$ is the one-dimensional velocity disperson. 
 Using the Equipartition theorem, total kinetic energy can be related to the temperature: $(mv^2 = {3}k_BT)$. 
 
  Putting equation \eqref{eq:growth_eqn} in $\nabla _{a }T^{a b }=0$ gives equation \eqref{eq:emt} in the limit of vanishing $ \mathcal{P}^{\langle ab \rangle} $.
 The SPT is based on the dust limit of the energy-momentum tensor, i.e $T^{ab}_{\rm{fluid}}=    u^au^b \rho_{m }  $, the EFTofLSS included the contribution of the pressure and anisotropic tensor. The EFTofLSS further assumed mode separability without justification because it is essential for the effective field approach. In sub-section \ref{sec:matter-horizon}, we will provide proof of separability.

\subsection{Separation of scales and astrophysical matter horizon}\label{sec:matter-horizon}

On large scales, where the characteristic size of a particle, $R_{\rm{ph}}$, is negligible compared to an external length scale of interest, $L_{\rm{ph}}$, i.e $R_{\rm{ph}}\ll L_{\rm{ph}}$,  the PPA yields a consistent approximation. 
However, it breaks down on non-linear scales or small scales where  $L_{\rm{ph}}\approx R_{\rm{ph}}$. 
The standard differential geometry approach to capturing the impact of local curvature on particle propagation is to calculate the critical point of the second variation of \eqref{eq:massive_particle_actionl} (i.e. the geodesic deviation equation) $ \frac{\d^2 \xi^{c}}{\d\tau^2} + R^{c}{}_{def} \xi^{d} u^{e} u^{f}  = 0 $, where $\xi^{c}$ is the deviation vector and $R^{c}{}_{def}$ is the Riemann tensor(see \cite{Umeh:2023lbc} for detials).  It determines whether two test particles which were initially moving parallel to each other would converge or diverge due to local curvature. We consider the limit where $\xi^a$ is Lie dragged along the integral curves of $u^a$: $\frac{{\d} \xi^{a}}{{\d} \tau}  = \nabla_{b} u^a \xi^b  $ and use the irreduciable covariant  decomposition  $\nabla_{b} u_{a} $  to split it into physical observables
\begin{eqnarray}\label{eq:decomposeCDU}
\nabla_{b} u_{a} &= -u^{b}A^{a}+\frac{1}{3}\Theta {h}^{ab }+\sigma^{ab} + \omega^{ab}.
\end{eqnarray}
where $A^a$ is the acceleration $A_a=u^d\nabla_{d} u_a$,  $\Theta ={\D}_{a}u^a$  describes the expansion/contraction of the nearby family of geodesics. It could be positive  $\Theta >0$ or negative $\Theta<0$, but the actual physical interpretation of $\Theta >0$  or  $\Theta <0$ depends on the orientation of the spacetime~\cite{Gaztanaga:2024vtr,Umeh:2026ajv}. 
  $\sigma_{ab} =h_{a}{}^{c}h_{b}^d \nabla_{\<c}u_{d\>}$ is the shear deformation tensor, which describes the rate of change of the deformation of nearby geodesics when compared to flat spacetime.  $\omega_{ab}=h_{a}{}^{c}h_{b}^d \nabla_{[c}u_{d]}$ is the vorticity tensor. %It is an anti-symmetric tensor. % It describes the twisting of nearby geodesics.  
The decomposition of the geodesic deviation equation in terms of these physical quantities leads to propagation equations $\Theta$,  $\sigma_{ab}$ and vorticity ${\omega}_{ab}$~\cite{Ellis:1998ct}. Without loss of generality, we provide the propagation equation ${\Theta}$ only
\begin{eqnarray}
\frac{ {\rm{D}} {{\Theta}} }{{\rm{D}} \tau} &=& - \frac{1}{3}{\Theta}^2 - {\sigma}_{ab}{\sigma}^{ab} 
 - {R}_{ab} {u}^a {u}^b\,,
\label{eq:expansion}
\end{eqnarray} 
where  ${ {\rm{D}} {{\cdots}} }/{{\rm{D}} \tau}  = u^a \nabla_{a} \cdots$ is the directional derivative and $R_{ab}$ is the Ricci tensor,

In a universe such as ours(almost FLRW), the expansion splits into global, $\Theta_{H}  $(Hubble flow)  and local parts, $\Theta_{L}$: $\Theta = \Theta_{H}  + \Theta_{L}$.
The local expansion ${\Theta_L} $ satisfies the following propagation equation~\cite{Umeh:2026ajv}
% which is coupled to the global expansion(we made use of the time-time component of GR to express ${R}_{ab} {u}^a {u}^b$ in terms of the matter density)
\begin{eqnarray}\label{eq:local-expansion}
\frac{ {\rm{D}} {{\Theta_L}} }{{\rm{D}} \tau} 
& =   - \frac{1}{3}{\Theta^{2}_L}-\frac{2}{3} \Theta_{H}\Theta_{L} - {\sigma}_{ab}{\sigma}^{ab} 
-\frac{1}{2}\kappa\left[  \delta \rho \right] \,,
\end{eqnarray}
where we made use of the time-time component of GR to express ${R}_{ab} {u}^a {u}^b$ in terms of the matter density, $\left[  \delta \rho \right] $ is the fluctuation of the matter density around the mean value. 
Since  ${\sigma}_{ab}{\sigma}^{ab} >0$ is positive definite, equation \eqref{eq:local-expansion} can be solved for over-dense regions $\delta \rho >0$ as partial differential inequality
 \begin{eqnarray}
\frac{1}{\Theta_{L}(\tau)}& \ge - \frac{1}{\exp\left[\mathcal{I}_{1} (\tau)\right]}  \left[ \frac{-1}{\Theta_{L\rm{ini}} }
+\mathcal{I}_{2} (\tau) \right]\,,
\label{eq:focusing_theorem2}
\end{eqnarray}
where $\mathcal{I}_{1} (\tau) $ is a function of the background expansion with
%\begin{eqnarray}
$\mathcal{I}_{1} (\tau) 
 \approx   -2 \ln (1+z)
 \,, $
 $
\mathcal{I}_{2} (\tau) 
\approx
-\frac{1}{3}\int_{z}^{\infty}\frac{ \d z}{(1+z)^3 H(z)}. $
% \qquad{\rm{for}}\qquad \tau \in \left[0,\tau_{\rm{max}}\right] \,
%\label{eq:I2}.
%\end{eqnarray}since $\Theta_{L}(\tau_{\star}, R_{\star})$
The terms in the square brackets vanish at finite time $\mathcal{I}_{2} (\tau) = 1/ {\Theta_{L\rm{ini}} }$, for converging initial data $\Theta_{L\rm{ini}} <0$ since $\mathcal{I}_{2} (\tau) <0$, this implies that at a finite time in the future, $\tau_{\star}$, the expansion vanishes $\Theta=0$ for a sub-region of finite extent~\cite{Umeh:2026ajv}. The family of geodesics within $r<R_{\star}$ cannot be extended beyond $\tau_{\star}$. This can easily be seen by evaluating an infinitesimal extension of the trajectory beyond $\tau_{\star}$: $\tau  = \tau_{\star}+ \Delta\tau $,  Implementing this to the volume element ${\rm{det}}\left[{\mathcal{J}}(\tau)\right] $ leads to 
\begin{eqnarray}
{\rm{det}}\left[{\mathcal{J}}(\tau_{\star})\right] \approx{\rm{det}}\left[
 {\mathcal{J}}(\tau)\right] \big[1-   \frac{1}{2}\left[ {\sigma}_{ab}{\sigma}^{ab} 
  + {R}_{ab} {u}^a {u}^b\right](\Delta \tau)^2\big]\,.
  \end{eqnarray}
  %Umeh:2026mac
This shows that if the weak energy condition holds $ R_{ab}  u^{a} u^{b}\ge0$, any infinitesimal extension of the geodesics leads to caustics ${\rm{det}}\left[{\mathcal{J}}(\tau)\right]\to 0$ in finite time.   

The
matter horizon $\Theta(\tau_{\star}, R_{\star}) = 0=\Theta_{H}(\tau_{\star})  + \Theta_{L}(\tau_{\star}, R_{\star})$, defines a unique proper time, $\tau_{\star}$, that is the proper time when a local sub-region with size $ r<R_{\star}$ decoupled from the Hubble flow. 
In GR, a consistent way of introducing a spatial length scale is via a proper length  
\begin{equation}
L = \int
\sqrt{
g_{ab} \frac{\mathrm{d}x^a}{\mathrm{d}\lambda}
\frac{\mathrm{d}x^b}{\mathrm{d}\lambda}
}
\mathrm{d}\lambda \,,
\end{equation} 
where $\lambda$ is an affine parameter. 
%We require that $r^{a} $ is orthogonal to $u^{a}$ so that the  proper length is given by
Without loss of generality, we require that the spacelike curve is geodesic:  $r^{a} \nabla_{b}r^{b} = 0$, where $r^{a} = \d x^{a}/\d \lambda$ is a spacelike 4-vector.
Similar to equation \eqref{eq:decomposeCDU}, the covariant decomposition of $\nabla_{a} r_{{b}} $ is given by
\begin{eqnarray}\label{eq:decomposera}
\nabla_{a} r_{{b}} 
=r_{a}\tilde{A}_{b} + \tfrac{1}{3}\gamma_{ab}\,\tilde\Theta 
+ \tilde\sigma_{ab} + \tilde\omega_{ab},
\end{eqnarray}
where $\gamma_{ab}$ is the metric on the timelike  hypersurface, 
 $\tilde\Theta \equiv \gamma^{ab}\nabla_a r_b$ is the expansion,
 $\tilde\sigma_{ab} = \tilde\sigma_{\langle ab\rangle} =  \gamma_{\<a}{}^c\gamma_{b\>}{}^d\nabla_c r_d$ is the symmetric tracefree shear,
$\tilde\omega_{ab}=\gamma_a{}^c\gamma_b{}^d\nabla_{[c}r_{d]}$ is the antisymmetric vorticity.  $\tilde{A}_b \equiv r^c\nabla_c r_b, $ is the acceleration of the congruence; it is orthogonal to $r^a$:  $\tilde{A}_b r^b=0$.  Using the Ricci identity, the propagation equations for $\tilde\Theta$, $\tilde\sigma_{ab} $ and $\tilde\omega_{ab}$ can be derived; they have a similar structure as propagation equations $\Theta$,  $\sigma_{ab}$ and vorticity ${\omega}_{ab}$ respectively. 
We consider the standard model of cosmology in conformal Newtonian gauge ~\cite{Umeh:2010pr}: 
\begin{eqnarray}\label{eq:metric}
\d  s^2  &= a^2\left[-(1 + 2\Phi)\d \eta^2 + \left(1-2 \Psi\right)\delta_{i j}  \d x^{i}\d x^{j}\right] \,,
\end{eqnarray}
where $\delta_{ij}$ is the spatial metric of the flat background spacetime,   $\Phi$ and $\Psi$ are scalar potentials.  We calculate $\tilde{\Theta}$  and express $\Phi$ and $\Psi$ in terms of the projected mass density, $\Sigma( r)$ using the Poisson equation
\begin{eqnarray}\label{eq:theta_Sigma}
\tilde{\Theta} (R) & \approx  \frac{3}{R}\left[1 +\frac{\sigma_{v}^2}{c^{2}}- \frac{1}{3 c^{2}}\int_{0}^{R} \d  r '{r'} \Sigma( r')\right]\,,
\end{eqnarray}
where $\bar{\tilde{\Theta}}  = 3/R$ ($R$ is the comoving distance in the sub-region and  $\sigma_{v}^2 = \<\left( v- \<v\>\right)\>^{2}$ is the velocity dispersion.  
$ \tilde{\Theta} (R)  $ vanishes at a finite proper  distance $R_{\star}$ where
$
\frac{1}{c^{2}} \int_{0}^{R_{\star}} \d  r' {r' }\Sigma( r')  = 3\left(1 +{\sigma_{v}^2}/{c^{2}}\right)
$.
  $\Sigma(R)$ is given in~\cite{Navarro:1995iw} for the NFW profile. Just as in the case of the timelike geodesics,  geodesics with initial condition at the centre of the sub-region cannot be extended beyond $R_{\star}$ without encountering caustics.

\section{Hierrachial multi-scale universe}\label{sec:Hierrachial}

\subsection{Matter horizon separatrix and piece-wise geodesics}

\begin{figure}[h]
\centering
\includegraphics[width=70mm,height=50mm] {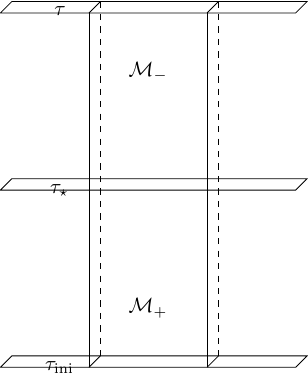}
\caption{A one-parameter family of timelike geodesics with initial condition set at $\tau_{\rm{ini}}$  in $({\cal M}^{+}, g_{ab}^{+})$ with local coordinate time flowing forward, it evolves into the future $\tau_{+} > \tau_{\rm{ini}}$ untill it reaches a maximal hypersurface at  $\tau_{\star}$.  At the maximal hypersurface, it decouples from $({\cal M}^{+}, g_{ab}^{+})$  and turns around (changes orientation) and continues to evolve into the future, $\tau_{-} >\tau_{\star}$ according to an observer at rest in $({\cal M}^{-}, g_{ab}^{-})$ manifold.   }
\label{fig:matching_spacetimes}
\end{figure}

In the standard model of structure formation, the dynamics of structure after collapse, it evolved as a separate FLRW universe. The formation of the matter horizon allows to build a more complete picture. 
In this set-up, the observed universe is described by a union of orientation-preserving manifolds $\mathcal{M} =  (\mathcal{M}_{+}\setminus \mathcal{D})\cup_{\phi} \mathcal{M}_{-} $.
The oriented manifolds on each side of the boundary are endowed with metrics $g_{ab}^\pm$, such that $({\cal M}^{+}, g_{ab}^{+})$ denotes a Lorentian manifold describing an epoch when the initial conditions for a family of nearby geodesics were set on an expanding background spacetime with coooredinate time flowing forward. 
We denote the Lorentian manifold  with the coordinate time orientation reversed as 
$({\cal M}^{-}, g_{ab}^{-})$. This is the manifold that the matter evolves on after decoupling from the Hubble flow. 
Both manifolds are time-oriented such that the geodesic initialised at $\tau_{\rm{ini}}$
evolves on  $({\cal M}^{+}, g_{ab}^{+})$  until it reaches a maximal hypersurface at $\tau_{\star}$ and decouples from the forward flowing coorinate time and continues its subsequent evolution on $({\cal M}^{-}, g_{ab}^{-})$  with the flow of cooredinate time reversed but with the proper time flowing forward with a discrete jump. at $\tau_{\star}$. This is illustrated in Figure \ref{fig:matching_spacetimes}.

\iffalse

 The  tangent vectors to the integral curves in $({\cal M}^{\pm}, g_{ab}^{\pm})$ are
\begin{eqnarray}
u^a_{+}
 =\frac{\d x^a_{+}}{\d \tau_{+}}~~~~~~~~~{\rm{and}}~~~~~~~ u_{-}^a     = \frac{\d x^a_{-}}{\d \tau_{-}}   \,,
\end{eqnarray}
where $x_{+}^{a}$ is associated with  the tangent space of ${\cal M}^{+}$ and ${x}^{a}_{-} $ is associated with the tangent space of ${\cal M}^{-}$. 
\fi
%Hence, the energy-momentum tensor in the particle limit decomposes as described in  Appendix \ref{sec:EMT} and also in \cite{Umeh:2023lbc}
The action of the  massive particle given in equation \eqref{eq:massive_particle_actionl} can be decomposed in a piece-wise fashion to apply to both manifolds with a boundary at the matter horizon
  \begin{eqnarray}\label{eq:piecewise_action}
%S (\gamma,\gamma') &=& S_{-} + S_{+} 
%\\
S (\gamma_{\pm},\gamma'_{\pm}) &=&
 \int_{\tau_{\rm{ini}}}^{\tau_{\star}}
 {L} _{+}\left[ \gamma_{+}( \tau_{+}) , \gamma'_{+}( \tau_{+})\right] \d  \tau_{+} 
 +\int_{\tau_{\star}}^{\tau_{\rm{final}}} 
 {L} _{-}\left[ \gamma_{-}( \tau_{-}) , \gamma'_{-}( \tau_{-})\right] \d  \tau_{-}
 \,,
\end{eqnarray}
where $L_{+}$ is the Lagrangian of the massive particle with initial conditions set on the expanding coordinates with forward flowing coordinate time, $L_{-}$ is the Lagrangian of the massive particle after decoupling from the forward flowing coordinate time.
  The critical point of equation \eqref{eq:piecewise_action} with respect to an infinitesimal variation, $s$,  corresponds to an infinitesimal variation of the respective actions
\begin{eqnarray}\nonumber
%0  = \frac{\d S }{\d s} \bigg|_{s = 0}&=& \frac{\d S _{-}}{\d s} \bigg|_{s = 0}+ \frac{\d S _{+}}{\d s} \bigg|_{s = 0}\\
 \frac{\d S }{\d s} \bigg|_{s = 0}&=&\frac{\d }{\d s} \bigg|_{s = 0} \int_{\tau_{\rm{ini}}}^{\tau_{\star}}
 L _{+}\left[\gamma_{+}( \tau_{+}),{{\gamma}_{+}}( \tau_{+})\right] \d  \tau_{+}
 + 
\frac{\d }{\d s} \bigg|_{s= 0} \int_{ \tau_{\star}}^{\tau_s}
 L _{-}\left[\gamma_{-}( \tau_{-}),{{\gamma}_{-}}( \tau_{-})\right] \d  \tau_{-}\,.
\label{eq:vary_Null_action}
\end{eqnarray}
\iffalse
Variation of both actions without implementing the proper variation condition gives 
\begin{eqnarray}
\frac{\d S_{-} }{\d s} \bigg|_{s = 0}&=& \frac{\partial L_{-}}{\partial {{\gamma'}^i_{-}} }( \tau)\xi^i_{{-}}( \tau) \bigg|_{\tau_{o}}^{\tau_{\star}}
+ \int^{ \tau_{\star}}_{\tau_s}  \left(\frac{\partial L_{-}}{\partial \gamma_{-}^i} - \frac{\d}{\d  \tau } \frac{\partial L_{-}}{\partial {{\gamma'}_{-}^i} }\right)\xi_{-}^i( \tau)  \d  \tau\ ,
\\
\frac{\d S_{+} }{\d s} \bigg|_{s = 0} &=& -  \frac{\partial L_{+}}{\partial {{\gamma'}_{+}^i} }( \tau)\xi_{+}^i( \tau)\bigg|_{\tau_{\star}}^{\tau_s}
+ \int^{\tau_s}_{ \tau_{\star}}  \left(\frac{\partial L_{+}}{\partial \gamma_{+}^i}- \frac{\d}{\d  \tau } \frac{\partial L_{+}}{\partial {{\gamma'}_{+}^i} }\right)\xi_{+}^i( \tau)  \d  \tau\,.
\end{eqnarray}
\fi
Performing the functional derivative of the Lagrangian and  imposing proper variation at the final  endpoints ($\xi^i(\tau_{\rm{ini}}) = \xi^i(\tau_{\rm{final}}) =0$) of the geodesic gives 
\begin{eqnarray}
0 &=& \left[ 
 \frac{\partial L _{+}}{\partial {{\gamma}_{+}^i} }(\tau_{\star})\xi_{+}^i( \tau_{\star}) - \frac{\partial L _{-}}{\partial {{\gamma}^i_{-}} }(\tau_{\star})\xi_{-}^i( \tau_{\star})  \right]
\\ \nonumber &&
+\int^{\tau_\star}_{\tau_{\rm{ini}}}  \left(\frac{\partial L_{+}}{\partial \gamma_{+}^i}( \tau_{+} )
- \frac{\d}{\d  \tau_{+}} \frac{\partial L_{+}}{\partial {{\gamma}_{+}^i} }(\tau_{+})\right)\xi_{+}^i( \tau_{+}) \d  \tau_{+} 
\\ \nonumber &&
+ \int^{\tau_{\rm{final}}}_{\tau_{\star}}  \left(\frac{\partial L_{-}}{\partial \gamma_{-}^i}( \tau_{-}) - \frac{\d}{\d  \tau_{-}} \frac{\partial L_{-}}{\partial {{\gamma}_{-}^i} } (\tau_{-}) \right)\xi_{-}^i(\tau_{-})  \d  \tau_{-}\,,
\end{eqnarray}
where $\xi^a_{\pm} $ is a deviation vector $\xi^a_{\pm} = {\partial x^a_{\pm}(\tau,s)}/{\partial s}\,. $
For a consistent variation, we require that the matter congruences are piece-wise smooth at the boundary, $\tau_{\star}$:
% Using equation \eqref{eq:discrte_map}, it becomes clear that  $\xi^i_{-}(\tau_{\star}) = -\xi_{+}^i (\tau_{\star})$. 
this  translates to the requirement that the Euler-Lagrange equations are independently  satisfied~\cite{Markvorsen:2022arXiv220713515M,Umeh:2023lbc}
\begin{eqnarray}\label{eq:EL_Null_A}
  \frac{\d}{\d  \tau_{+} } \frac{\partial L_{+}}{\partial {{\gamma'}_{+}^i} }-\frac{\partial L_{+}}{\partial \gamma_{+}^i} &=&  0 
\qquad {\rm{for}}\quad   \tau_{+} \in [\tau_{\rm{ini}}, \tau_{\star}] \,,
 \\
   \frac{\d}{\d  \tau_{-}} \frac{\partial L_{-}}{\partial {{\gamma'}_{-}^i} } -\frac{\partial L_{-}}{\partial \gamma_{-}^i}&=&  0 
  \qquad {\rm{for}} \quad  \tau_{-} \in [\tau_{\star},\tau_{\rm{final}}] \,.
  \label{eq:EL_Null_C}
\end{eqnarray}
And at the boundary we have % the following condition that must hold $\xi^i_{+}( \tau_{\star})  = \xi^i_{-}( \tau_{\star})  =\xi^i( \tau_{\star}) $
\begin{eqnarray}\label{eq:Israel}
 \left[     \frac{\partial L_{+}}{\partial {{\gamma}_{+}^i} }(\tau_{\star}) +  \frac{\partial L_{-}}{\partial {{\gamma}_{-}^i} }(\tau_{\star}) \right]\xi^i_{+}( \tau_{\star})  = 0\,.
\end{eqnarray}
Considering the massive particle Lagrangian (i.e equation \eqref{eq:massive_particle_actionl})
leads to geodesic equations in both sectors $
u_{+}^a \nabla_a u^{b}_{+} = 0\,,$ and  $ u_{-}^a \nabla_a u^{b}_{-} = 0$
and  the boundary conditions for the geodesics $u^a_{+} \big|_{\mathcal{N}}+u^a_{-}\big|_{\mathcal{N}} = 0$. Following the splitting of the action (equation \eqref{eq:vary_Null_action}), the  energy-momentum tensor splits as well
 \begin{eqnarray}%\label{eq:EMT_particles2}
 T^{ab}_{+} &=& \sum^N_{\ell}  T^{ab}_{+ \ell}
 = \sum^N_{\ell}  \frac{m_{+\ell}}{\sqrt{-g_{+}}} \int_{\rm{ini}}^{\star} {\d}\tau_{+}\, u^{a}_{+\ell} u^{b}_{+\ell} \delta^4(x_{+} - x_{+\ell}(\tau_{+\ell})) \,,
\\ 
 T^{ab}_{-} &=& \sum^N_{\ell}  T^{ab}_{- \ell}
 = \sum^N_{\ell}  \frac{m_{-\ell}}{\sqrt{-g_{-}}} \int_{\star}^{\rm{final}} {\d}\tau_{-}\, u^{a}_{-\ell} u^{b}_{-\ell} \delta^4(x_{-} - x_{-\ell}(\tau_{-\ell})) \,.
\end{eqnarray}

\subsection{Einstein-Hilbert action on a manifold with boundary}

\begin{figure}[h]
\centering 
\includegraphics[width=100mm,height=55mm] {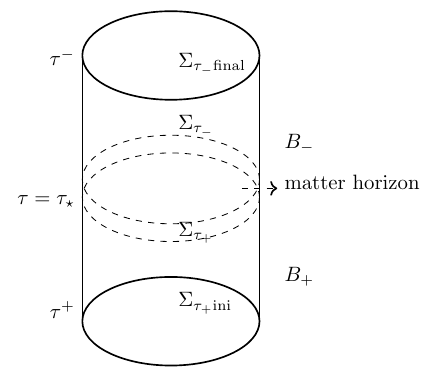}
\label{fig:spacetime_diagram_short}
\caption{
We illustrate the matching of spacetimes at a common hypersurface.
The timelike boundaries $B_\pm$ enclose the spatial region, while the spacelike boundaries denote where the initial data is defined.  The boundary at $\tau_{\star}$ hosts gravitational edge modes.
}
\end{figure}

We can now extend the same formalism to the full Einstein field equations. For this case, we consider
  figure \ref{fig:spacetime_diagram_short} for visualise guridance. Note that we focus on particles in the overdense regions since they will surely decouple from the Hubble flow at a finite time in the future. 
The projected metrics on the two sheets of spacetime are related according to $h^{-}_{ab} = \Omega^2_{S} h^{+}_{ab}$, where $\Omega_{S} \equiv \Omega_{-}/\Omega_{+}$ is the ratio of effective scale factors. Considering scalar perturbations on hypersurfaces of constant proper radius, the induced metrics $\gamma^{\pm}_{ab} = g^{\pm}_{ab} - r^{\pm}_{a}r^{\pm}_{b}$ are related according to $\gamma^{-}_{ab} = \Omega^2_{T} \gamma^{+}_{ab}$, with $\Omega_{T} \equiv a_{-}/a_{+}$. For details on the derivation of these results, see \cite{Umeh:2026mac}. The standard Israel-Darmois conditions do not cover this configuration; hence, we find that the diffeomorphism generating vector field  $X^a(x)$ satisfies the conformal Killing equation \cite{Umeh:2026ajv} with the solution 
\begin{eqnarray}\label{eq:isometry}
{X}^{b}_{\Sigma/B}= \alpha^{b} + M^{b}{}_{a} x^{a}_{+} + \lambda x^{b}_{+} + 2 (x_{+a} \beta^{a}) x^{b}_{+} - (x_{+a} x^{a}_{+}) \beta^{b},
\end{eqnarray}
where the constant parameters $\{\alpha^a, M^a{}_b, \lambda, \beta^b\}$ correspond to translations, rotations, dilatations, and special conformal transformations, respectively. On $\Sigma$ hypersurface, this forms an $SO(4,1)$ group, which is the isometry group of de Sitter space, while on $B$-hypersurface, it is an $SO(3,2)$ group, which is the isometry group of Anti-de Sitter space. This immediately shows that the matter horizon breaks the diffeomorphism group on the hypersurface $\rm{Diff}(h)$ to the isometry group.  The Goldstone modes associcated with the breaking of the gauge symmetry is what it refered to as the gravitational edge mode~\cite{Donnelly:2020xgu,Donnelly:2022kfs}.

The  action of the gravitational theory on ambient spacetime $\mathcal{M} = ( \mathcal{M}_{+}\setminus \mathcal{D}) \cup \mathcal{M}_{-}$ is given by
\begin{eqnarray}\label{eq:gravity_action}
S_{\rm{Full}}\left[g_{ab}\right] &= S_g\left[g^{+}_{ab}\right]  + S_g\left[g^{-}_{ab}\right] + S_{\mathrm{bd}}\left[h^{{\pm}}_{ab},\gamma^{\pm}_{ab}, N^{\pm}_{ab}\right]  \,,
\end{eqnarray}
where  $S_g\left[g^{\pm}_{ab}\right] $ is the sum of Einstein-Hilbert, $ S_{\mathrm{EH}}$ and matter fields $S_{\mathrm{M}} $ actions:
$
S_g\left[g^{\pm}_{ab}\right]  = S_{\mathrm{EH}} \left[g_{ab}^{\pm}\right] +S_{\mathrm{M}}\left[g_{ab}^{\pm}\right ]
$:
\begin{eqnarray}
 S_{\mathrm{EH}} \left[g^{\pm}_{ab}\right]   &=&  \frac{1}{2\kappa}
\int_{\mathcal{M}}\mathcal{R}\left[g^{\pm}_{ab}\right]  \,\sqrt{-g_{\pm}}\d^4 x _{\pm} \,,
  \end{eqnarray}
  where $\mathcal{R}$ is the Ricci scalar and $S_{\mathrm{M}}  \propto \delta^4(x^a-x_{\ell} (\tau_{\ell}))S (\gamma_{\pm},\gamma'_{\pm})$ is the action of the matter field with $S$ given in equation \eqref{eq:piecewise_action}.  The steps on how to vary the action in the ambient spacetime in the presense of a boundary are given \cite{Umeh:2026ajv,Umeh:2026mac}. 
  % the essential point is that the energy-momentum tensor decomposes into $ T^{ab}_{+} (\tau_{\star})$ and  $ T^{ab}_{-}$.
\begin{eqnarray}\nonumber
\delta S_g\left[g_{ab}\right] & =&
\int_{\mathcal{M}^{+}}\frac{1}{2}\left(\frac{1}{\kappa}G^{+}_{ab}+\Lambda g_{ab}-T^{+}_{ab}\right)\delta g^{ab}_{+}\,\sqrt{-g_{+}} d^4x 
 \\ \nonumber& &
+\int_{\mathcal{M}^{-}}\frac{1}{2}\left(\frac{1}{\kappa}G^{-}_{ab}+\Lambda g_{ab}-T^{-}_{ab}\right)\delta g^{ab}_{-}\,\sqrt{-g_{-}} d^4x 
 \\ & &
 +\frac{1}{2\kappa}  \int_\mathcal{M_{+}}  \nabla_{a} \delta V^a_{+} \sqrt{-g_{+}} d^4x_{+}  +\frac{1}{2\kappa}  \int_\mathcal{M_{-}}  \nabla_{a} \delta V^a_{-} \sqrt{-g_{-}} d^4x_{-}\,,
 \label{eq:totalaction_variation}
%   +\frac{1}{2\kappa} \oint _{\partial {\mathcal {M}}} %k_{b} \left[\delta V^b_{-}\d\Sigma^{-}_{b}  + \delta V^b_{+} \d\Sigma^{+}_{b}\right]
\end{eqnarray}
where $G^{\pm}_{ab} =  R^{\pm}_{ab} - {1 \over 2} g^{\pm}_{ab} \mathcal{R}^{\pm}$ is the Einstein tensor,  $\delta V^c_{\pm} $ are boundary terms resulting from the variation of the Ricci tensors associated with both manifolds $\mathcal{M}_{\pm}$: $g^{ab}_{\pm} \delta R^{\pm}_{ab} = \nabla_{a} \delta V^a_{\pm}$. It is given by $ \delta V^c_{\pm} =  \left[g^{ab}_{\pm} \delta  \Gamma^c_{\pm ab} - g^{a c}_{\pm} \delta \Gamma^b_{\pm ab}\right]$. Again $T_{ab}^{\pm}$ is the respective energy-momentum tensors for the standard matter(e.g. baryons)
$
T^{{\pm}}_{ab} \equiv -\frac{2}{\sqrt{-g_{\pm}}}
 \frac{\delta S^{\pm}_{\rm{M}}}{\delta g^{\pm}_{ab}}  \,.
$ 
The pre-symplectic potential $\phi_{\pm} =  u^{\pm}_a (\delta V^a_{\pm}) \sqrt{-h_{\pm}} +  r^{\pm}_a (\delta V^a_{\pm}) \sqrt{-\gamma_{\pm}} $  can be decomposed into conjugate momenta and boundary terms
 \begin{eqnarray}\nonumber
    \frac{1}{2\kappa} \int_{\mathcal{M}} \nabla_a (\delta V^a_{\pm}) \sqrt{-g_{\pm}}\, d^4x_{\pm} &=&
\int _{ { {\Sigma}}_{\pm}} \sqrt{h^{\pm}}\bigg\{
  -   \frac{1}{\kappa} \delta u^0_{\pm}  K^{\pm}
+ \bigg[\Pi^{\pm}_{ab}
+L^{\pm}_{(a} u^{\pm}_{b)} \bigg]\delta h^{ab}_{\pm}\bigg\} \d^3{x}_{\pm} 
\\ \nonumber &&
+\int _{{B}}  \sqrt{\gamma^{\pm}}\bigg\{
  \frac{1}{\kappa}\delta r^r_{\pm}  \tilde{K}^{\pm}
+ \bigg[\tilde{\Pi}^{\pm}_{ab}
+\tilde{L}^{\pm}_{(a} r^{\pm}_{b)} \bigg]\delta \gamma^{ab}_{\pm}\bigg\} \d^3{x}_{\pm}
\\ \nonumber &&
-
 \frac{1}{\kappa} \int _{{ {\Sigma}}_{\pm}}
   \delta\bigg[\sqrt{h^{\pm}} K^{\pm}\bigg] \d^3{x_{\pm}}
-
 \frac{1}{\kappa} \int _{{B}_{\pm}}
   \delta\bigg[\sqrt{\gamma^{\pm}} \tilde{K}^{\pm}\bigg] \d^3{x_{\pm}}
   \\ &&
   +\frac{1}{\kappa}  \oint _{{\partial B}_{\pm}}\sqrt{N^{\pm}} \delta\left( r^a_{\pm} u_{\pm a}\right) \d^2{x}_{\pm} \,.
   \label{eq:decomposed_boundary_term}
\end{eqnarray}
where $ K^{\pm}$ is the trace of the extrinsic curvature tensor of the spacelike hypersurface.  
We have introduced the covariant conjugate momenta:
\begin{eqnarray}
\Pi_{ab} &=&- \frac{2}{\sqrt{h}}\frac{\delta I_{g}}{ \delta h^{ab}} = \frac{1}{\kappa} \bigg[  {K} h_{ab} -K_{ab} \bigg]\,, \qquad
L_{a}  =  -\frac{2}{\sqrt{h}}\frac{\delta I_{g}}{ \delta u^{a}}  =  \frac{1}{\kappa}  \bigg[{K} _{ab}  u^{b}\bigg] \,,
 \\
\tilde{\Pi}_{ab} &=&- \frac{2}{\sqrt{\gamma}}\frac{\delta I_{g}}{ \delta \gamma^{ab}} = -\frac{1}{\kappa} \bigg[  \tilde{K}\gamma_{ab} -\tilde{K}_{ab} \bigg] \,,
\qquad 
\tilde{L}_{a} = -\frac{2}{\sqrt{\gamma}}\frac{\delta I_{g}}{ \delta r^{a}}  =  -\frac{1}{\kappa}  \bigg[\tilde{K}_{ab}  r^{b} \bigg]\,.
\end{eqnarray}
The pre-symplectic term  from the variation of the Einstein-Hilbert action needs regularisation; that is, the standard boundary terms must be added \cite{Umeh:2026mac}
%\begin{eqnarray}\nonumber
$ S_{\mathrm{bd}}\left[h^{{\pm}}_{ab},\gamma^{\pm}_{ab}, N^{\pm}_{ab}\right]  =  S_{\mathrm{GHY}}\left[h^{{\pm}}_{ab}\right] + S_{\mathrm{GHY}}\left[ \gamma^{{\pm}}_{ab}\right]  
+  S_{\mathrm{Hayward}}\left[N^{{\pm}}_{ab}\right] \,,$
%\end{eqnarray}
where  $ S_{\mathrm{GHY}}\left[h^{{\pm}}_{ab}\right]$  and $S_{\mathrm{GHY}}\left[ \gamma^{{\pm}}_{ab}\right] $ are the Gibbon-Hawking-York boundary term on the spacelike and timelike hypersuface respectively and $ S_{\mathrm{Hayward}}$ is the Hayward corner term, it depends on the metric on the screen space, $N_{ab}$~\cite{Gibbons:1976ue,Brown:1992br,Hayward:1993my}. 
Using the relationship between the variation of the projected metric tensors and the full spacetime tensor
\begin{eqnarray}\label{eq:variationhab}
  \delta   h_{\pm}^{ab} = \delta g_{\pm}^{ab} +\delta u_{\pm}^{a}u_{\pm}^{b} + u_{\pm}^{a}\delta u_{\pm}^{b}\,. \qquad  \delta   \gamma_{\pm}^{ab} = \delta g_{\pm}^{ab} -\delta r_{\pm}^{a}r_{\pm}^{b} - r_{\pm}^{a}\delta r_{\pm}^{b}\,,
\end{eqnarray}
we de-project some of the terms to the bulk
   \begin{eqnarray}\nonumber
\int _{ {\Sigma^{\pm}}}Z^{\pm}_{ab} \delta h^{ab}_{\pm}\sqrt{h^{\pm}} \d^3 x_{\pm}&=&
   \int_{\mathcal{M}^{\pm}}\,\delta\left( \tau_{\pm}(x^{\pm}) - \tau_{\star}\right)
     Z^{\pm}_{ab}\delta g_{\pm}^{ab}   \sqrt{-g_{\pm}} d^4x_{\pm}  
    % \\  \nonumber &&
     +    \int _{ {\Sigma^{\pm}}}\sqrt{h^{\pm}} L^{\pm}_{b}\delta u_{\pm}^{b}  \d^3 x_{\pm}   \,.
     ~~~~~
\end{eqnarray}
and for the timelike hypersurface, we have 
\begin{eqnarray}\nonumber
\int _{{B^{\pm}}} \sqrt{\gamma^{\pm}} 
\tilde{Z}^{\pm}_{ab} \delta \gamma_{{\pm}}^{ab} d^3 x_{\pm}  
 &=&   \int_{\mathcal{M}^{{\pm}}}\,\delta\left( R_{\pm}(r_{\pm}) - R_{\star}\right)
      \tilde{Z}^{\pm}_{ab} \delta g_{{\pm}}^{ab}      \sqrt{-g_{_{\pm}}} d^4x_{\pm}
        %   \\  \nonumber &&
     +    \int _{ {B^{\pm}}}\sqrt{\gamma^{\pm}}  \tilde{L}^{\pm}_{b}\delta r_{\pm}^{b}   \d^3 x_{\pm}\,.
 ~~~
\end{eqnarray}
Putting all these together and imposing a consistent variational principle, i.e ${\delta S_{\rm{Full}}\left[ g_{ab}\right] }/{\delta g_{ab}}   = 0$, we impose piece-wise continuity at the boundary region and require that, which leads to the following equations of motion~\cite{Umeh:2026mac}
\begin{eqnarray}\label{eq:EoM}
G_{+}^{ab}+\Lambda g_{+}^{ab} = {\kappa} \tau_{+}^{ab} \,,\qquad G_{-}^{ab} +\Lambda g_{-}^{ab}= {\kappa} \tau_{-}^{ab} \,,
\end{eqnarray}
and the boundary energy flux condition $
\sum_{{\ell=1}}^{N}\left[\mathcal{L}_{\ell a}^{+}  -\mathcal{L}_{\ell a}^{-}\right] = 0, $ $
\sum_{\ell =1}^{N}\left[\tilde{\mathcal{L}}_{\ell a}^{+}  -\tilde{\mathcal{L}}_{\ell a}^{-} \right] = 0\,.
$. Note that $K$ vanishes at the boundary. 
 $\tau_{ab}$ is the effective energy-momentum tensor~\cite{Umeh:2026mac}
\begin{eqnarray}\label{eq:effective_emt}
 \tau^{\pm}_{ab} 
      &\approx \sum_{\ell=1}^{N}\bigg[\rho_{m \ell}{u}^a_{\pm \ell} {u}^b_{\pm \ell}  
     + \delta\left( R(x_{\pm}) - R_{\ell \star}\right)       \tilde{Z}^{\pm}_{\ell ab}  \bigg] \,.
 \end{eqnarray}
where $\rho_{m  \ell}=  { m^{\pm}_{\ell}}\delta^{3}\left(x^i_{\pm} - \gamma^i_{\pm \ell}(t_{\pm})\right) /{\sqrt{h^{\pm}}}$ is the standard baryon matter density and
$\tilde{ Z}^{\pm}_{ab} $ is a geometric backreaction contribution; it is a direct physical consequence of "stitching" two scales together. %the macro-scale universe to the micro-scale galaxy.
 It is given by $\tilde{ Z}^{\pm}_{ab} = \tilde{ \Pi}^{\pm}_{ab}+2\tilde{L}^{\pm}_{(b} r^{\pm}_{a)}$,
where  $\tilde{\mathcal{L}}_{ a}^{+}  = \sqrt{\gamma} \tilde{L}_{a}$ is the momentum flux along the timelike boundary and   $\tilde{\Pi}^{\pm}_{ab}$ is the canonical momentum conjugate to the induced metric $\gamma_{ab}$. 
The contribution to $ \tau^{\pm}_{ab} $ from the spacelike boundary is subdominant~\cite{Umeh:2026mac}. 
% It is sometimes referred to as the Brown–York quasilocal stress tensor~\cite{York:1972sj}.
 $\tilde{\Pi}^{\pm}_{ab} $ and $\tilde{L}_{a}$ can be expressed in terms of the extrinsic curvature tensor~~\cite{Brown:1992br}:
$
\tilde{\Pi}^{\pm}_{ab} 
= - \big[  \tilde{K}\gamma^{\pm}_{ab}
-\tilde{K}_{ab} 
\big] /{\kappa} $
 and $
\tilde{L}^{\pm}_{a}= - \big[\tilde{K}_{ab}  r^{b}_{\pm} \big]/{\kappa} $.
In order to interpret $\tilde{Z}^{ab}_{\pm}$ as part of the effective energy-momenton tensor,  we  decompose it with respect to ${u}^a_{{\pm}} $
$
\tilde{Z}^{ab}_{\pm} = \tilde{\rho}_{{\pm}} {u}^a_{{\pm}} {u}^b_{{\pm}} + \tilde{P}_{{\pm}}{h}^{ab}_{{\pm}}
  +2 \tilde{q}^{(a}_{{\pm}} {u}^{b)}_{{\pm}} + \tilde{\pi}^{\<ab\>}_{{\pm}},
$
where $\tilde{ \rho}_{{\pm}} $, $\tilde{P}_{{\pm}}$, $\tilde{q}_{{\pm} a}$ and $\tilde{\pi}^{\<ab\>}_{{\pm}}$ are the boundary energy density,  pressure,   energy flux vector and anisotropic stress tensor respectively.  Again, these are the gravitational edge modes, they are Goldstone modes resulting from the breaking of the diffeomorphism group down to the isometry group at the boundary~\cite{Donnelly:2020xgu,Takayanagi:2019tvn}.  
%As we showed in equation \eqref{eq:isometry}, the diffeomorphism group is broken down to the isometry group at the matter horizon. 
Note that $\tilde{K}_{ab}^{\pm} =  \gamma^{c}_{\pm}{}_{a}{\nabla}_{c} r^{\pm}_{b}$, so using equation \eqref{eq:decomposera}, we find that $ \tilde{Z}^{\pm}_{ab}  = \tilde{\sigma}^{\pm}_{\<ab\>}/{\kappa}  $.
We focus on the  energy density and  pressure; the full expression can be found in \cite{Umeh:2026mac}
\begin{eqnarray}
\tilde{\rho}_{{\pm}}  &=\tilde{Z}_{ \pm ab}{u}^a_{\pm} {u}^b_{\pm} = \frac{1}{\kappa} \big[ {u}^a_{\pm} {u}^b_{\pm} \tilde{\sigma}_{\<ab\>} \big]\,,
\quad \tilde{P}_{\pm}  = \frac{1}{3}\tilde{\rho}_{{\pm}} 
\label{eq:geometric_matter}
\end{eqnarray}
Equation \eqref{eq:effective_emt} gives the total microscopic contributions to the energy-momentum tensor labelled by particle position, $\gamma^{i}$ and the matter horizon or physical size of the particle. 

However, we are interested in the effective energy-momentum tensor at a single time scale (see figure \ref{fig:obs_multi_scale}). For example, dynamics in the Hubble flow, $ \tau^{+}_{ab}$ is given in equation \eqref{eq:effective_emt}, which is a sum over the individual energy-momentum tensors of clusters of galaxies, while $ \tau^{-}_{ab}$  is the sum over the energy-momentum tensors of galaxies that make up one cluster of galaxy.  This setup is general; it can apply to any time scale captured in figure \ref{fig:obs_multi_scale}  provided the metric tensor has a conformal Minkowski form (equation \ref{eq:metric}). 
Tracking the dynamics of each of the particles  could be very challenging, but for a large number of them, we can replace the sum with an average just as we did in the standard cosmology limit(equation \eqref{eq:standardEMT}):
   \begin{eqnarray}\label{eq:fluidEqn}
 \tau^{ab}_{\pm\rm{fluid}}
&\equiv & \frac{1}{ V_{\pm}}   \frac{1}{ R_{\pm}} \int_{\Sigma_{\pm}}\int_{R_{\pm}}   \tau^{ab}_{\pm} \sqrt{h_{\pm}} \, \sqrt{h^{\pm}_{RR}}  d^3\gamma_{\pm} \d r_{\star}\,, ~~~~~~
 \end{eqnarray}
 where $\gamma^{i}$ is the coordinate position of a particle, $R = \int h_{RR} \d r_{\star}$ is the matter horizon and $V_{\pm} = \int \d^{3 } \gamma_{\pm}\sqrt{h_{\pm}}$.  After some straightforward algebra, we find
 \begin{eqnarray}\label{eq:total_emt_{dust}}
 \tau^{ab}_{\rm{fluid}\pm} =
 \rho^{\pm}_{T}  {u}^a_{\pm} {u}^b_{\pm} + {P}_{ T\pm} {h}^{ab}_{\pm}   +2 {q}^{(a}_{T \pm} {u}^{b)}_{\pm} + {\pi}^{\<ab\>}_{T\pm} \,,
\end{eqnarray}
where $ \rho_{T} =  \rho^{\pm}_{ m} +\hat{\rho}_{\pm}$(sum of standard matter density, $\rho_{m} = M/V$ and backreaction contribution $\hat{\rho}_{\pm}$, $ {P}_{ T\pm}  =  \hat{P}_{ \pm}  +\mathcal{P}_{\pm}  + \hat{\mathcal{P}}_{\pm}+\mathcal{S}_{\pm} $, $ {q}^{a}_{T \pm}  =  \hat{q}^{a}_{ \pm} $ and ${\pi}^{\<ab\>}_{T\pm}  = \hat{\pi}^{\<ab\>}_{\pm} + \mathcal{P}^{\langle ab \rangle}_{\pm}  + \hat{\mathcal{P}}^{\langle ab \rangle}_{\pm}  +\mathcal{S}^{\langle ab \rangle}_{\pm} $. 
The additional contributions to the pressure and anisotropic stress tensor are due to thermal velocities associated with $ \rho^{\pm}_{ m} $, $\hat{\rho}_{\pm}$ and $ \hat{P}_{ \pm} $~\cite{Umeh:2026mac}. Also, we introduced the bulk viscosity term, $\zeta_{\pm}$, which describes the resistance to uniform expansion or collapse $\hat{P}_{ \pm}  = \frac{1}{3} \zeta_{\pm} \hat{\rho}_{ \pm} $  and shear viscosity, $\eta$, which describes the resistance to shape deformations
$ \hat{\pi}^{\<ab\>}_{\pm}  = \eta_{\pm} \sum_{{\ell = 1}}^{N}  \hat{\sigma}^{\<ab\>}$ (see \cite{Eckart:1940te} for details).
$ \tau_{\rm{fluid}\pm}^{ab} $ satisfies the conservation equation  in a piece-wise limit
$\nabla^{a}  \tau_{\rm{fluid} + }^{ab}  = 0$ and $\nabla^{a}  \tau_{\rm{fluid} -}^{ab}  = 0$,  for diffeomorphisms Lie-dragged along the integral curves of the matter field.
The components of the conservation equation in the limit $\hat{\Pi}^{cb}_{T\pm}  =0= \hat{q}^a_{T\pm}$ is given by \cite{Umeh:2026mac}
\begin{eqnarray}
 &\dot{\rho}_{T\pm}  +\left({\rho}_{T\pm}+{P}_{T\pm}\right) \left( \bar{\Theta}_{\pm} 
 + {\D}_a v^a_{\pm}\right)   = 0\,,
   \label{eq:fluid_limit_energy_conservation_eqn}
 \\
 &\dot{v}_a + \frac{1}{3}  \left( \bar{\Theta}_{\pm} 
 + {\D}_{b} v^{b}_{\pm}\right) v_{\pm a} + {\D}^a \Phi_{\pm}   + \frac{{\D}_a {P}_{T\pm}}{(\rho_{\pm} + {P}_{T\pm})} =0\,.
    \label{eq:fluid_limit_momentum_conservation_eqn}
\end{eqnarray}
Here, $v^a_{\pm}$ is the relative velocity between the matter and comoving frames. Eqs. \eqref{eq:fluid_limit_energy_conservation_eqn}–\eqref{eq:fluid_limit_momentum_conservation_eqn} differ from the standard dust result only by backreaction contributions to energy density and pressure (see \eqref{eq:geometric_matter}). 

\section{Galaxy flat rotation curves}
We now show how the backreaction terms lead to diversity in rotation curves for galaxies in various stages of evolution as described in figure \ref{fig:obs_multi_scale}.
For purely azimuthal motion, $v^a_{\pm}= (0,0, v^{\phi}_{\pm})$, the steady-state limit of equation \eqref{eq:fluid_limit_momentum_conservation_eqn} yields the rotation velocity.
  \begin{equation}\label{eq:rotationvelo}
v_{\phi}
=
\begin{cases}
\sqrt{\,r_{-}\,\frac{\d\Phi_{-}}{dr_{-}}  +\mathcal{Z}_{-}
\frac{\d \ln \hat{\rho}_{-}}{\d \ln r_{-}}  +  \mathcal{Y}_{-}
\frac{\d \ln {\rho}_{m -}}{\d \ln r_{-}} 
}
& r_{-} < r_\star, \qquad %\Delta r <0 
\\ %[10pt]
\sqrt{\,r_{+}\,\frac{\d\Phi_{+}}{dr_{+}}  +\mathcal{Z}_{+}
\frac{\d \ln \hat{\rho}_{+}}{\d \ln r_{+}}  +  \mathcal{Y}_{+}
\frac{\d \ln {\rho}_{m+}}{\d \ln r_{+}} 
 }
& r_{+} \ge r_\star  \qquad  %\Delta r >0.
\,.
\end{cases}
\end{equation}
The linearity of the Poisson equation, $\nabla^2 \Phi \approx 4\pi G(\rho_{m} + \hat{\rho})$, implies that the gravitational potential can be decomposed $\Phi_{\pm}(r) = \Phi_{m}^{\pm}(r) + \hat{\Phi}_{\pm}(r)$, where $\Phi_{m}$ represents the baryonic potential and $\hat{\Phi}$ the contribution from backreaction.
 The first integral of the baryonic Poisson equation gives 
 \begin{equation}
 \frac{\d\Phi_{m\pm}}{\d r_{\pm}} =\frac{r_{\rm{ini}}^2}{r^2_{\pm}}\frac{\d\Phi_{\pm-}}{\d r_{\pm}} \big|_{r_{\pm}=r_{\rm{ini}}}  +\frac{G M_{bk \pm}(r_{-}< r_{\star})}{r^2_{\pm}} ,
 \end{equation} 
where we adopted the Hernquist density profile  to calculate $M_{bk \pm}$ \cite{Hernquist:1990ApJ} 
 \begin{equation}
 \rho_{ m \pm }(r_{\pm}, a_{\pm})  =  \frac{M_{\pm}}{2\pi} \frac{a_{\pm}}{r_{\pm}(r_{\pm} + a_{\pm})^3}\,,
 \end{equation} 
 $"a_{\pm}"$  is a free scale parameter and $M_{\pm}$ is the total mass.  For a spherical mass distribution, we set the inner boundary condition $\frac{d\Phi_{m-}}{dr_{-}}|_{r_{\rm{ini}}} = 0$, while ensuring flux continuity at the boundary layer as per equation \eqref{eq:EoM}.

Calculating $\d \hat{\Phi}_{\pm}/\d r_{\pm}$ from its corresponding Poisson equation $\nabla^2 \hat{\Phi}_{\pm} = 4\pi G \hat{\rho}_{\pm}$ is more involved because $\hat{\rho}_{\pm}$ given in equation \eqref{eq:geometric_matter} is related
$\tilde\sigma_{\pm ab} $, which satistifes the following propagation equation \cite{Umeh:2026mac}
\begin{equation}\label{eq:sigmaeqn}
r^c_{\pm}\nabla_c \tilde\sigma_{\pm ab} 
+\tfrac{2}{3}\tilde\Theta\,\tilde\sigma_{\pm ab}
+ \mathcal{E}_{ab}
- \tfrac{1}{2}\mathcal{R}_{ab}=   0\,.
\end{equation}
Given equation\eqref{eq:metric}, the solution to equation \eqref{eq:sigmaeqn} is given by 
\begin{equation}
 {u}^a_{\pm} {u}^b_{\pm} \tilde\sigma_{\pm ab}\approx  \frac{1}{r^2_{\pm}} \int_{r_{\pm\rm{ini}}}^{r} {r'_{\pm}}^2 \gamma^{ab}_{\pm} \left[{ \D}_{a}{\D}_{b} \Phi_{\pm} \right]\d r_{\pm}\,.
 \end{equation}
Using equation \eqref{eq:fluidEqn}, the bulk backreaction density becomes
\begin{eqnarray}\label{eq:energy_densitymicro}
\hat{\rho}_{\pm} &=&\frac{1}{R_{\pm}} \sum_{\ell = 1}^{N}   \tilde{\rho}_{\pm \ell} 
\xrightarrow{N \to \infty} \; \frac{\langle \tilde{\rho}_{\pm} \rangle}{R_{\pm}}
=\frac{1}{\kappa}
\frac{2}{r_{\star}r^2_{\pm}}Q_{\pm} (r_{\pm})\,,
\end{eqnarray}
where we replaced the sum with an all-sky average, and $ R  = a\int \d r_{\star} = r_{\star} $ and  the parameter $Q_{\pm}$ is given by
$
Q_{\pm} (r_{\pm}) = 
 \left[ r_{\pm} \Phi_{\pm}(r_{\pm}) \right]^{r}_{r_{\rm{in}}} - \int_{r_{\pm\rm{ini}}}^{r_{\pm,}} \Phi_{\pm}(r_{\pm}) \d r'_{\pm}\,.
$
Using the Poisson equation, we found that $\hat{\Phi}_{\pm}$ satisfies an integro-differential equation
 \begin{eqnarray}\label{eq:intrgro-diff}
 \frac{\d^2  \hat{\Phi}_{\pm}  }{\d r^2_{\pm}} + \frac{2}{r_{\pm}} \frac{d \hat{\Phi}_{\pm}  }{dr_{\pm}}
 &=  
 \frac{1}{r^2_{\pm} r_{\star}}Q_{\pm} (r_{\pm}) \,.
 \end{eqnarray}
 By setting $g_{\pm}  (r_{\pm}):=r^{2}_{\pm}\d \hat{\Phi}_{\pm}/\d r_{\pm}$ it becomes clear that homogenous limit of equation \eqref{eq:intrgro-diff} is a Modified Bessel equation of order 1 and the source term is given by ${r_{\pm}}{\Phi_{m\pm}'(r_{\pm})/r_{\star}}$. Therefore, the general solution becomes
 \begin{eqnarray}\label{eq:backreactionAccen}
      \frac{d\hat{\Phi}_{\pm}}{dr_{\pm}} &=   \frac{1}{{r^{2}_{\pm}}}
\Bigl[A_{\pm}g_{1\pm}(r_{\pm}) +B_{\pm}\,g_{2\pm}(r_{\pm}) 
       + g_{p\pm}(r_{\pm}) \Bigr]\,.
     \end{eqnarray}
where $g_{1\pm}$ and $g_{2\pm}$ are two linearly independent solutions to the homogeous equation
 \begin{eqnarray}
g_{1\pm}(r_{\pm}) = 2\sqrt{\frac{r_{\pm}}{r_{\star}}} I_1\left( 2\sqrt{\frac{r_{\pm}}{r_{\star}}} \right), 
\quad g_{2\pm}(r) =2\sqrt{\frac{r_{\pm}}{r_{\star}}}K_1\left( 2\sqrt{\frac{r_{\pm}}{r_{\star}}} \right)
   \end{eqnarray}
and the particular solution
\begin{eqnarray}
g_{p\pm }(r_{\pm}) &= \frac{GM_{\pm}}{2}\Bigl[ g_{1\pm}(r_{\pm})\int_{r_{\rm{ini}} }^{r_{\pm}} \frac{g_{2\pm}(r'_{\pm})\,r'}{(r'_{\pm}+a_{\pm})^{2}}\,dr'_{\pm} ~~~~
%\\ \nonumber  &
 g_{2\pm}(r_{\pm})\int_{r_{\rm{ini}} }^{r_{\pm}}  \frac{g_{1\pm}(r'_{\pm})\,r'_{\pm}}{(r'_{\pm}+a_{\pm})^{2}}\,dr'_{\pm} \Bigr].
\end{eqnarray}
  The gravitational potential is obtained by integrating equation \eqref{eq:backreactionAccen}
     \begin{eqnarray}
\hat{\Phi}_{\pm}&= \hat{\Phi}_{\pm 0}+ 
A_{\pm} \mathcal{E}_{1\pm}(r_{\pm}) +B_{\pm}\,\mathcal{E}_{2\pm}(r_{\pm}) +\mathcal{E}_{p\pm}(r_{\pm}) 
~~~~
\end{eqnarray}
where $\mathcal{E}_{1\pm}$, $\mathcal{E}_{2\pm}$ and $\mathcal{E}_{p\pm}$ are integrals over $g_{1\pm}$, $g_{2\pm}$ and $g_{p\pm}$ respectively.  In general,  $\hat{\Phi}_{0\pm}$ is determined in terms of the two arbitrary constants $A_{\pm}$ and $B_{\pm}$, however, in our case, equation \eqref{eq:intrgro-diff} is independent of $\hat{\Phi}_{0\pm}$  at $r = r_{\rm{ini}}$. 
Therefore, we determine it independently by imposing the physical condition consistent with that of baryons.
We require that the $\hat{\Phi}_{-}$ is regular at  $r_{\rm{ini}} $, thus, 
$\hat{\Phi}_{-}(r_{\rm{ini}})  =  0 $,  $
\frac{d\hat{\Phi}_{-}}{dr_{-}} \big|_{r =r_{\rm{ini}} } =0$, hence, $B_{-}$ must vanish since $K_{1}$ diverges in the limit $r\to 0$ leading to $\hat{\Phi}_{- 0} = 0$ and  $A_{-} = {- g_{p-}(r_{\text{ini}}) {r_{\star}}{} /2 r_{\text{ini}}}$.
For the exterior region, we impose the continuity condition at $r = r_{\star}$: $\hat{\Phi}_-(r_{\star}) = \hat{\Phi}_+(r_{\star})  $ and $
  \frac{d{\hat{\Phi}}_{-}}{dr_{-}} \big|_{r_{\star}} = \frac{d\hat{\Phi}_+}{dr_{+}} \big|_{r_{\star}} $. There are two possible classes of galaxy rotation curves depending on the evolutionary stage of the galaxy. This is illustrated in figure \ref{fig:obs_multi_scale}, we consider each case below:
       \begin{figure}[h]
\centering 
\includegraphics[width=100mm,height=60mm] {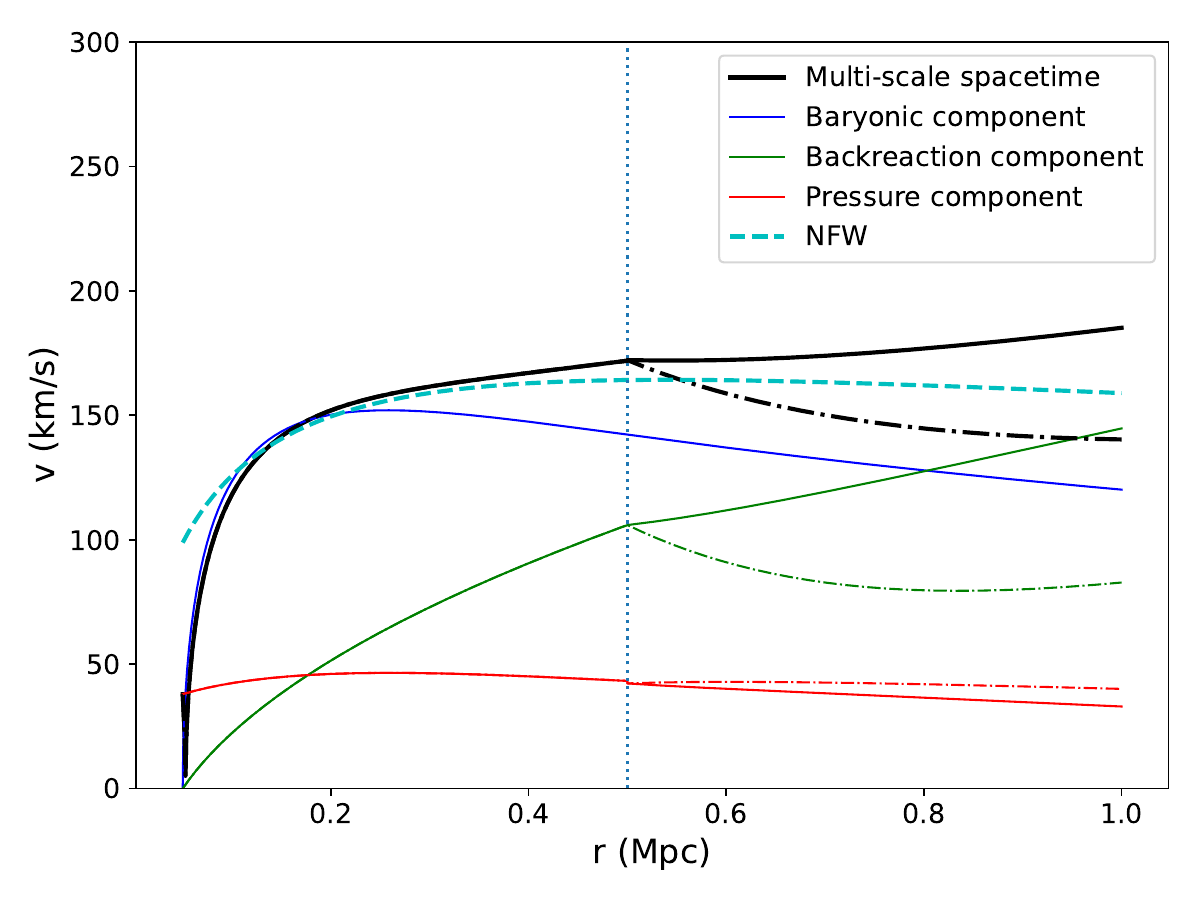}
\label{fig:rotation_curve}
\caption{The galaxy rotation curves of a typical galaxy with  Hernquist density profile for the baryon density with parameters set to  $a_{\pm} = 0.2$ Mpc, and $ M =1.5\times10^{12} M_{\otimes}$. 
The velocity dispersion is set to $\sigma_{T\hat{\rho}_{\pm}1D}^2= \sigma_{\hat{\rho}_{\pm}1D}^2 = 20 $[km/s].
The NFW prediction is added for comparison~\cite{Navarro:1995iw}.
}
\end{figure}
  \begin{itemize}
   \item 
  $\tau_{{\rm{gal}}}$-hypersurface, the exterior region is given by a spacetime with boundary at infinity,, hence $A_{+}$ must vanish since $I_{1}$ grows rapidly  $I_1(y) \sim \frac{e^y}{\sqrt{2\pi y}}$ as $y\to \infty$, theerefore, $\hat{\Phi}_{+ 0}  = \hat{\Phi}_-(r_{\star}) $ and  $B_{+}  =  \frac{r_{\star}^2 }{2 K_1(2)}  \frac{d{\hat{\Phi}}_{-}}{dr_{-}} \big|_{r_{\star}}  -\frac{ g_{p-}(r_{\star})}{2 K_1(2)}$.
  
\item 
$\tau_{{\rm{clus}}}$-hypersurface, the exterior region is given by a spacetime with a finite extent at the galaxy cluster boundary, hence, the general solution can be approximated with the growing component leading to $\hat{\Phi}_{+ 0}  = \hat{\Phi}_-(r_{\star}) $ and $A_{+}  =  \frac{r_{\star}^2 }{2 I_1(2)}  \frac{d{\hat{\Phi}}_{-}}{dr_{-}} \big|_{r_{\star}}  -\frac{ g_{p-}(r_{\star})}{2 I_1(2)}$.
  \end{itemize}

  Furthermore, $ \mathcal{Z}_{\pm} $ and  $ \mathcal{Y}_{\pm}$ are functions of disperson velocity and galaxy bias \cite{Umeh:2026mac} $ \mathcal{Z}_{\pm}  =  \mathcal{Z}_{\pm}(\sigma_{T\hat{\rho}_{\pm}1D}^2 , \hat{\rho}_{\pm}/\rho_{m\pm})$ and  $  \mathcal{Y}_{\pm} = \mathcal{Y}_{\pm} = \mathcal{Z}_{\pm}(\sigma_{T\hat{\rho}_{\pm}1D}^2 , \hat{\rho}_{\pm}/\rho_{m\pm})$.  The galaxy rotation curves obtained from solving equation \ref{eq:intrgro-diff} are given in figure \ref{fig:obs_multi_scale}, it gives both the limits of rotational curves observed in dwarfs and massive galaxies~\cite{OBrien:2017ogc}. The exactly flat rotation curve may be obtained by relaxing the isothermal approximation.
  
  Finally, the total Newtonian gravitational force (${a}_N$) (sum of the baryon component and the backreaction component) displays MOND-like feature \cite{1983ApJ...270..365M}:
       \begin{eqnarray}\label{eq:MOND}
 {a}_N & =& \frac{d\Phi_{\pm}}{dr_{\pm}} = \frac{G M_{bk\pm}}{r_{\pm}^{2}} \left[ 1 + \nu_{\pm}(r_{\pm}) \right]\,,
 % \frac{G M_{bk \pm}}{r^2_{\pm}} + \frac{2 A_{\pm}}{r^2_{\pm}} \sqrt{\frac{r_{\pm}}{r_{\star}}} I_1\left( 2\sqrt{\frac{r_{\pm}}{r_{\star}}} \right)\,,
       \end{eqnarray}
  where   $\nu_{\pm}(r_{\pm}) = \left[ A_{\pm}\, g_{1\pm}(r_{\pm}) + g_{p\pm}(r_{\pm}) \right]/{G M_{bk\pm}} $. In the Deep-MOND regime, it scales like $C/r_{\pm}$ largely independent of the particular solution for the galaxy in $\tau_{{\rm{clus}}}$ evolutionary phase. % or the form of the source term in equation \eqref{eq:intrgro-diff}.

\section{Conclusions}\label{sec:conc}

The challenge of long dynamical range has long hindered the precise modelling of matter distribution in the universe. In this paper, we have made several key contributions that resolve this bottleneck while forging a novel connection between gravitational edge modes and dark matter phenomenology.

First, we identified a fundamental feature of general relativity: geodesics defining the flow of matter on spacetime can cease to be geodesics at finite time or spatial extent, with breakdown preceded by a matter horizon. This provides a physically well-defined criterion for separating scales in cosmological structure formation.

Second, by systematically identifying matter horizons, we described how the full spacetime can be partitioned into a hierarchy of domains or sub-regions related by discrete transformations at shared boundaries. Glueing these sub-regions via manifold surgery anchored on the variational principle yields a geometric backreaction effect on particle trajectories that is absent in the standard point-particle treatment.

Third, we established that this covariant backreaction effect corresponds precisely to what is known in quantum gravity as gravitational edge modes; Goldstone modes resulting from the breaking of the diffeomorphism group down to the isometry group at finite boundaries. These edge modes contribute physical degrees of freedom that encode additional gravitational energy not captured by the local bulk stress-energy tensor of standard matter.

Fourth, we derived the effective energy-momentum tensor incorporating these edge mode contributions, showing that they enter as additional density, pressure, and viscosity terms. The conservation equations were obtained, revealing how edge modes modify the dynamics of gravitational collapse.

Fifth, we applied this framework to galaxy rotation curves, demonstrating that gravitational edge modes naturally produce the observed flattening in galactic outskirts. We derived analytic expressions for the rotation velocity (equation \eqref{eq:backreactionAccen}) and showed that the effective Newtonian force displays MOND-like features (equation \eqref{eq:MOND}) in the deep-MOND regime, all without invoking dark matter particles.

Finally, our framework provides a first-principles, multi-scale description of matter clustering at any resolution, resolving the singularity issues inherent in the standard point-particle approximation. It offers a concrete physical interpretation of the effective dark matter required by observations—not as exotic particles, but as gravitational edge modes arising from the fundamental structure of spacetime itself.

\section*{Acknowledgement}
I benefited immensely from discussions with Sravan Kumar. I appreciate the support of the CIC Foundation; without them, this work would not have seen the light of day. 
The computations in this paper were done with the help of tensor algebra software xPand \cite{Pitrou:2013hga}, which is based on xPert~\cite{Brizuela:2008ra}.

%\bibliographystyle{../JHEP}%{JHEP}%
%\bibliography{$HOME/Dropbox/UWC_papers/q-dipole/draft/cosmoref.bib}

\providecommand{\href}[2]{#2}\begingroup\raggedright\endgroup

\end{document}